\documentstyle[prl,aps,twocolumn,axodraw]{revtex}
\newenvironment{namelist}[1]{%
\begin{list}{}
	{
	\settowidth{\labelwidth}{#1}
	\setlength{\leftmargin}{1.1\labelwidth}}}{%
\end{list}}
\begin{document}
\draft
%
%
\title{
\Large\bf
Field-Theoretic Realization of Heavy Meson as Composite Particle
}
\author
{{\bf 
Chi-Yee Cheung and Wei-Min Zhang
}\\
{\it 
Institute of Physics, Academia Sinica, Taipei 11529, Taiwan, 
Republic of China}
\\}
\date{\today}
\maketitle
\begin{abstract}

We construct a realistic field-theoretic model for the structure of a
heavy meson in the heavy quark limit.  The model is fully covariant and
satisfies heavy quark symmetry.  The Isgur-Wise function, the decay
constant, and the axial vector coupling constant are studied.  This model
overcomes the limitations caused by non-covariance of the light-front
quark model, and provides an ideal framework to systematically evaluate
the heavy quark symmetry breaking effects caused by the $1/m_Q$ correction
terms in the heavy quark effective theory of quantum chromodynamics. 

\vskip 0.25 true cm 
\noindent PACS numbers: 14.40, 12.38, 12.40.A 
\end{abstract} 
%
\def\beq{\begin{equation}}
\def\eeq{\end{equation}}
\def\bea{\begin{eqnarray}}
\def\eea{\end{eqnarray}}
\def\cL{{\cal L}}
\def\Phiv{\Phi_v}
\def\Phivd{\Phi_v^\dagger}
\def\hv{h_v}
\def\hvb{\bar h_v}
\def\psiv{\psi_v}
\def\psivb{\bar\psi_v}
\def\psiq{\psi_q}
\def\psiqb{\bar\psi_q}
\def\qb{\bar q}
\def\mM{m_M}
\def\mQ{m_Q}
\def\mq{m_q}
\def\del{\partial}
\def\Lb{\bar\Lambda}
\def\vdot{v \cdot}
\def\vpdot{v'\cdot}
\def\GamM{\Gamma_M}
\def\GamMp{\Gamma_{M'}}
\def\sh #1 {#1\!\!/}
\def\vsh{v\!\!\!/}
\def\psh{p\!\!\!/}
\def\esh{\epsilon\!\!/}
\def\gam5{\gamma_5}
\def\gammu{\gamma_\mu}
\def\Gammu{\Gamma_\mu}
\def\pmu{p_\mu}
\def\delmu{\partial_\mu}
\def\gammuu{\gamma^\mu}
\def\Gammuu{\Gamma^\mu}
\def\pmuu{p^\mu}
\def\delmuu{\partial^\mu}
\def\ie{i\epsilon}
\def\d4p{{d^4p\over (2\pi)^4}}
\def\dlf{{dp^+d^2p_\bot\over (2\pi)^3 2p^+}}
%
%
\vskip 0.25 true cm

It is well known that in the infinite quark mass limit, quark-gluon
interaction becomes flavor and spin independent, leading to considerable
simplification in the description of heavy hadron
physics\cite{IW,Neubert}.  However, even in this hypothetical limit, low
energy quark-gluon dynamics is nonperturbative, and the problem of
calculating heavy hadron bound states directly from QCD remains
unsolved\cite{Zhang}.  Consequently, in order to quantitatively study 
heavy hadron physics, one still has to model the structures of the
hadrons phenomenologically. 
In order to study the transition properties of hadrons at arbitrary
momentum transfers, it is essential to have a fully Lorentz covariant
model for the bound state structure of hadrons, otherwise one often runs
into ambiguities and may obtain incorrect results. 
Unfortunately, none of the available quark models meets this requirement.

Among the various quark models, the light-front quark model comes 
close to being truely relativistic.  However, the fact that it is not 
fully covariant\cite{CCH,CCHZ} severely limits its usefulness.  
In a previous paper\cite{CCHZ}, we studied a covariant light-front 
model of heavy mesons, which provided a partial solution to the 
Lorentz covariance problem.  However, the approach taken there is not 
systematic, and light-quark currents are not considered.
In this work, we propose to construct a realistic heavy meson bound 
state in a field theoretic approach, so that
covariance is automatically guaranteed.  
This model satisfies heavy quark symmetry (HQS), and 
at the same time has the simplicity of the quark model picture. 
Calculations of hadronic matrix elements in this model reduce to 
computing Feynman diagrams which are finite.  
With this model, we can systematically and
quantitatively study the HQS breaking effects caused by the $1/\mQ$
correction terms in the heavy quark effective theory 
(HQET)\cite{EG} of quantum chromodynamics (QCD), 
which is important for a thorough understanding of heavy hadron physics.

In this letter we will present the basic formalism, and study the 
Isgur-Wise function, the decay constant, and the axial vector coupling 
constant.  
Furthermore, connection will be established with our previous work 
on covariant light-front model\cite{CCHZ}. 
Evaluation of HQS breaking effects will however be left to a forthcoming 
longer paper. 

To begin, we represent a pseudoscalar heavy meson effectively by a quantum 
field operator $\Phi$, with the familiar free Lagrangian  
\beq
\cL_0^M = \delmu\Phi^\dagger \delmuu\Phi
-\mM^2\Phi^\dagger\Phi,
\eeq
where $\mM$ is the meson mass.  In order to be consistent with HQET, 
we remove the heavy quark mass $\mQ$ from $\cL_0^M$ 
by redefining $\Phi$ as
\beq
\Phi(x) = {1\over \sqrt{\mM}} ~e^{-i\mQ \vdot x} ~\Phiv (x),
\label{Phiv}
\eeq
where $v$ is the velocity of the heavy meson (or its constituent  
heavy quark). In terms of the new field $\Phiv$, $\cL_0^M$ becomes 
$(\mQ \rightarrow \infty)$  
\beq
\cL_0^M=2 \Phivd (i\vdot\stackrel{\leftrightarrow}{\del}-\Lb)\Phiv
\eeq
where $\Lb=\mM-\mQ$, and 
$\stackrel{\leftrightarrow}{\del}={1\over 2} 
(\stackrel{\rightarrow}{\del}-\stackrel{\leftarrow}{\del})$.  
Thus $\Phiv$ corresponds to a particle with mass $\Lb$. 

The pseudoscalar heavy meson ($\Phiv$) couples to the heavy ($\psi_Q$) 
and light ($\psiq$) quarks via  
\beq
\cL_I^{M} = -G_0 \Phiv \hvb i\gam5 F(-i\vdot\del)\psiq
               + h.c., \label{LIM}
\eeq
where $G_0$ is the coupling constant, 
\beq
\hv(x) = {1+\vsh\over 2} e^{i\mQ\vdot x} ~\psi_Q(x)
\eeq
is the familiar reduced heavy quark field operator in HQET, and 
$F$ is a vertex structure function related to the heavy meson bound state 
wave function, on which we shall impose the constraint that the 
heavy meson does not decay into $Q$ and $\qb$ physically.
Note that $F$ is independent of the residual momentum of 
the heavy quark so that the heavy quark charge is not spreaded out, 
otherwise the Isgur-Wise function will not be properly normalized.

To study the compositeness of the the heavy meson, we shall assume that,
after summing the spin-flavor independent part of the interaction
between a heavy quark and gluons in HQET, the effective coupling between
the heavy quark and a light quark can be written as
\beq
\cL_I^{Qq} = g_0 \hvb i\gam5 f(-i\vdot\del)\psiq \cdot
                    f(i\vdot\del) \psiqb i\gam5 \hv
\eeq
in the pseudoscalar channel, 
where $g_0$ is the coupling constant, and $f$ is a form factor, 
whose presence is expected for an effective interaction resulting from 
non-perturbative QCD dynamics.  
$\cL_I^{Qq}$ can be considered as a generalized 
four-fermion coupling model\cite{NJL} inspired by QCD in the heavy 
quark limit. 

Connection between the meson picture and the quark picture is 
established by demanding that $\cL_I^{M}$ and $\cL_I^{Qq}$ give the same 
heavy-light quark scattering amplitude as depicted in Fig. 1.  
In the meson-quark interaction picture, 
the renormalized scattering amplitude is given by 
\beq
A_M = {i G^2 F(\vdot p) F(\vdot p')
      \over 2(\vdot k -\Lb)-G^2 \Pi_F^r(\vdot k)},
\label{AM2}
\eeq
where $\Pi_F(\vdot k)$ is the meson self-energy 
from the heavy-light quark loop shown in Fig. 1, which has been expanded 
around the mass shell $\vdot k=\Lb$: 
\beq
\Pi_F(\vdot k) = \Pi_F(\Lb) + \Pi_F'(\Lb) (\vdot k - \Lb)
               + \Pi_F^r(\vdot k), 
\label{PiF}
\eeq
and the renormalized quantities are defined by 
\bea
\Lb &=& \Lb_0 + {1\over 2} G_0^2 \Pi_F(\Lb),\\
G &=& \sqrt{Z_3} G_0,\\
Z_3 &=& 1+{1\over 2} G^2 \Pi'_F(\Lb)\label{Z3}.
\eea
The corresponding scattering amplitude in the quark-quark 
interaction picture is,
\beq
A_{Qq} = g_0 f(\vdot p') f(\vdot p) 
         {i\over 1-g_0\Pi_f(\vdot k)},
\label{AQq}
\eeq 
where $\Pi_f$ is the same as $\Pi_F$, except that the structure function 
$F$ is replaced by $f$.
If we demand that the interaction is strong enough to produce a 
bound state of mass $\Lb$,
then $A_{Qq}$ should have a pole at $\vdot k=\Lb$, which 
implies that $g_0 = {1/\Pi_f(\Lb)}$, and
\beq
A_{Qq}={-i f(\vdot p') f(\vdot p) \over \Pi'_f(\Lb) (\vdot k-\Lb)+
       \Pi_f^r(\vdot k)},
\label{AQq2}
\eeq
where $\Pi_f(\vdot k)$ has been expanded as in Eq. (\ref{PiF}).
From Eqs. (\ref{AM2}) and (\ref{AQq2}), 
it is seen that for $A_M=A_{Qq}$, we must have
$F=f$, so that $\Pi_F=\Pi_f=\Pi$, and 
\beq
G^2 = {-2/ \Pi'(\Lb)}.
\label{G}
\eeq
Hence the compositeness condition for the heavy meson fixes the strength 
of the ($\Phiv Qq$)-coupling vertex through Eq. (\ref{G}).
As we shall see later, Eq. (\ref{G}) is related to the wave function
normalization condition in the light-front quark model. 
The above discussion can be readily generalized to include 
heavy vector mesons ($\Phiv^\mu$); by heavy quark spin symmetry,
the vertex structure function $F$ must be the same for pseudoscalar and 
vector heavy mesons.  Due to the lack of space, we will skip 
the details here.

We have now constructed an effective relativistic quantum field theory 
for quarks and heavy mesons.  
The complete Lagrangian is given by
\beq
\cL = \cL^M_0 +\cL^M_I + \cL^q_{QCD} + \cL^{1/m_Q}_{HQET},
\eeq
where the notations are self evident.
Within this framework, hadronic matrix elements are calculated 
via standard Feynman diagrams.  
The Feynman rules are the same as in QCD and HQET, 
except the meson-quark vertex, for which the Feynman rule is 
\beq
(\Phiv Qq){\rm -vertex} = -i G F(\vdot p) \GamM,
\eeq
where $p$ is the momentum of the light quark, and 
$\GamM = i\gam5 (-\esh)$ for $M$ = pseudoscalar (vector) meson.
This model is simpler to work with than the ordinary light-front quark 
model, moreover it can be used to calculate hadronic form factors at 
arbitrary momentum transfers.

In the limit $\mQ\rightarrow\infty$, flavor symmetry is trivially 
satisfied in our model.  To demonstrate that spin symmetry also holds, 
we calculate the transition matrix element between heavy mesons 
due to the external heavy quark current (see Fig. 2a) 
\beq
J_\mu(x) = \bar h_{v'}^{Q'}(x) \Gamma_\mu \hv^Q(x), \qquad
(\Gammu=\gammu~{\rm or}~\gammu\gam5),
\eeq
\bea
\langle &M&'_{Q'q}(v')|J_\mu|M_{Qq}(v)\rangle 
= -i G^2\int\d4p F(\vpdot p) F(\vdot p) \nonumber\\
&& \cdot {Tr\left[(-\psh+\mq)\GamMp {1+\vsh'\over 2}\Gammu 
   {1+\vsh\over 2}\GamM \right]
   \over (\Lb-\vpdot p+\ie)(\Lb-\vdot p+\ie)(p^2-\mq^2+\ie)},
\label{MJM}
\eea
where $|M(v)\rangle$ stands for a heavy meson 
(pseudoscalar or vector) state, with the normalization
$\langle M(v')|M(v)\rangle = (2\pi)^3 2v^0 \delta^3(\Lb v' - \Lb v)$.
By Lorentz covariance, %
\beq
\psh\rightarrow a\vsh+b\vsh'
\eeq
with
\beq
a = {\vdot p-\vpdot p \vdot v' \over 1-(\vdot v')^2},\
b = {\vpdot p-\vdot p \vdot v' \over 1-(\vdot v')^2}.
\eeq
Moreover, since $\vdot\epsilon=\vpdot\epsilon'=0$, and 
$\vsh\gam5=-\gam5\vsh$, Eq. (\ref{MJM}) can be rewritten as
\bea
\langle &M&'_{Q'q}(v')|J_\mu|M_{Qq}(v)\rangle \nonumber\\
&=& -\xi(\vdot v')~
Tr\left[{1\over 4}\GamMp (1+\vsh')\Gamma (1+\vsh)\GamM\right],
\eea
where
\bea
\xi(\vdot v') && = iG^2\int\d4p F(\vpdot p) F(\vdot p) \nonumber\\
&&\cdot {(a+b+\mq) \over (\Lb-\vpdot p+\ie)(\Lb-\vdot p+\ie)(p^2-\mq^2+\ie)}
\label{xi}
\eea
is called the Isgur-Wise function.  Thus we have proved that 
in the $\mQ\rightarrow\infty$ limit, transitions between heavy mesons 
due to external heavy quark currents are described by only one independent 
form factor $\xi(\vdot v')$.  

Similarly, the heavy meson self-energy $\Pi(\vdot k)$ can be easily 
calculated, which in turn yields
\beq
G^{-2} = i\int \d4p {F^2(\vdot p) (\vdot p +\mq) 
         \over (\Lb-\vdot p+\ie)^2(p^2-\mq^2+\ie)}. 
\label{Gm2}
\eeq
through Eq. (\ref{G}).  
Thus we see that when $v=v'$, $\xi(1)=1$, as required by HQS.

For the heavy meson decay constant, we calculate the 
Feynman diagram shown in Fig. 2b.  The result is
\beq
\langle 0|\psiqb\Gammu\hv^Q|M_{Qq}(v)\rangle 
= F_M ~Tr\left[{1\over4} \Gammu (1+\vsh)\GamM\right],
\eeq
where $\Gammu=\gammu$ or $\gammu\gam5$, and 
\beq
F_M = i 2 \sqrt{3} G \int\d4p {F(\vdot p) (\vdot p + \mq)
\over (\Lb-\vdot p+\ie)(p^2-\mq^2+\ie)}
\label{FM}
\eeq
is the decay constant in the heavy quark limit, which is the same for 
pseudoscalar and vector heavy meson.  
$F_M$ is related to the usual meson decay constant $f_M$ by 
$F_M=\sqrt{\mM} f_M$.  

Finally, we consider the axial vector coupling constant ($g$) by evaluating 
matrix elements of the axial vector current
$A_\mu^a = \psiqb {\lambda^a\over 2} \gammu\gam5\psiq$.
$g$ is related through PCAC to 
the strength of interactions between heavy mesons and Goldstone bosons 
($\phi^a$)\cite{Yan}. 
In Ref. \cite{CCHZ}, we have shown how to conform with covariance in 
calculating $\xi(\vdot v')$ and $F_M$ in a light-front quark model, 
however doing the same with $g$,
which involves a purely light-quark current, remains a problem unsolved.
With the field-theoretic approach proposed in the present work, Lorentz
covariance is automatically guaranteed for heavy-quark and light-quark 
currents alike.
Hence the axial vector coupling constant extracted in this model 
should be more reliable.  The calculation is by now straightforward, 
and the answer is (see Fig. 2c)
\beq
\langle M'_{Qq'}(v)|A^a_\mu|M_{Qq}(v)\rangle 
= g ~Tr\left[ {1\over 4}\gammu\gam5 \GamMp (1+\vsh) \GamM\right],
\eeq
where we have omitted the SU(3) matrix element 
$\chi_{M'}^\dagger \lambda^a \chi_M$, and 
the axial vector coupling constant $g$ is given by 
\bea
g =&& - i{G^2\over 3} \int \d4p F^2(\vdot p)\nonumber\\
   &&\cdot {p^2 + 3\mq^2 + 6\mq\vdot p +2(\vdot p)^2 
    \over (\Lb-\vdot p+\ie)(p^2-\mq^2+\ie)^2}.
\label{acc}
\eea

To explicitly evaluate $\xi(\vdot v')$, $F_M$, and $g$, we need to 
specify the structure function $F$, which is unfortunately not calculable
at present.  Nevertheless, from the constraints that  
(1) $F$ does not depend on the heavy quark momentum and (2) it forbids 
on-shell dissociation of the heavy meson to $Q\qb$, a plausible 
form for $F$ is 
\beq 
F(\vdot p) = \varphi(\vdot p) (\Lb-\vdot p),
\label{FF}
\eeq
where the function $\varphi$ does not have a pole at $\vdot p = \Lb$.
The integrations over $p^0$ or $p^-$ in Eqs. (\ref{xi}), 
(\ref{Gm2}, (\ref{FM}), and (\ref{acc}) can be easily performed.  
To facilitate comparison with light-front quark model results, 
we carry out the $dp^-$-integrations, and obtain  
\bea
\xi(\vdot v')=G^2 \int &&\dlf \varphi(\vpdot p)\varphi(\vdot p)\nonumber\\
    &&\cdot (a+b+\mq)
\label{xi2}
\eea
\beq
F_M=2\sqrt{3} G \int \dlf \varphi(\vdot p)
    {(\vdot p+\mq)},
\label{FM2}
\eeq
\bea
g = -{G^2\over 3} \int && \dlf ~{1\over p^+} {\del\over\del p^-}
\{\varphi^2(\vdot p)~(\Lb-\vdot p)\nonumber\\
&&\cdot[p^2+3\mq^2+6\mq\vdot p+2(\vdot p)^2]\},
\label{acc2}
\eea
and
\beq
G^{-2} = \int \dlf \varphi^2(\vdot p) (\vdot p+\mq),
\eeq
where $p^+\ge 0$ and $p^2=\mq^2$. 

It is interesting to observe that, if  
\beq
\varphi(\vdot p) = {e^{-\vdot p/\omega}\over \sqrt{\vdot p+\mq}},
\label{phi}
\eeq
then $\xi(\vdot v')$ and $F_M$ are the same as those obtained in the 
covariant light-front quark model\cite{CCHZ}, 
and $G$ equals the wave function normalization constant.
We note that $e^{-\vdot p/\omega}$ is just the covariant light-front wave 
function proposed in \cite{CCHZ}, and the factor $\sqrt{\vdot p+\mq}$
is originated from the Melosh transformation. 

However $e^{-\vdot p}$ is not bounded when $p$ is off the mass shell, 
so that we must use $e^{-|\vdot p|}$ in a four dimensional integral,
which, together with the square root sign in (\ref{phi}), 
is repugnant in a field theoretic formalism.  
Hence we will choose instead the more well behaved form, 
\beq
\varphi(\vdot p)={e^{-(\vdot p)^2/2\omega^2}\over \vdot p+\mq-\ie},
\eeq
which yields very reasonable results both in the heavy quark limit and 
for $1/\mQ$ corrections . 
If we take $F_M\simeq F_B$, $f_B=0.18$ GeV\cite{Bernard}, 
and $\mQ=0.25$ GeV, then Eq. (\ref{FM2}) gives $\omega = 0.60$ GeV.
The resulting Isgur-Wise function is almost indistinguishable from that 
of Ref. \cite{CCHZ}.  
HQET analyses of inclusive semileptonic $B$ and $D$ 
decays find $\Lb \simeq 0.45$ GeV\cite{Falk}, which implies that 
\beq
g = 0.32
\label{acc3}
\eeq
in the heavy quark limit.  
Eq. (\ref{acc3}) is consistent with the constraint of
$g_{D^*D\pi} < 0.7$ obtained from $D^*$-meson decay width\cite{ACCMOR};
it also compares well with the QCD sum rules results of 
$g=0.21 \sim 0.39$ \cite{Belyaev}.   
Clearly, in order to make more precise physical predictions, 
we must also take into account $1/\mQ$ corrections, which will be 
treated in another publication.

We have now completed the specification of our field-theoretic model for
the bound state structure of a heavy meson. The heavy
meson bound state we have constructed, which obeys HQS, can be considered
as one due to the lowest order Hamiltonian in the HQET.  Hence it provides
an appropriate basis for studying systematically the HQS breaking $1/\mQ$
effects.  For example, Fig. 3 shows an $1/\mQ$ correction to the
Isgur-Wise function $\xi(\vdot v')$ in this model, which can be 
easily evaluated.   

In summary, we have developed a realistic field-theoretic model for heavy
meson as a composite particle.  Lorentz covariance allows us to extract
hadronic transition form factors at arbitrary momentum transfers without
running into ambiguities; hence it overcomes a severe drawback caused by
non-covariance in the light-front quark model.  Our model yields
physically realistic results, and provides an ideal framework in which
$1/\mQ$ corrections to HQS can be evaluated quantitatively and
systematically.  Results for $1/\mQ$ effects and further extention of the
model will be published elsewhere.

\acknowledgments
This work is supported in part by the National Research Council of the 
Republic of China under Grant Nos. NSC87-2112-M-001-002 
and NSC86-2816-M-001-008L.


\newcommand{\bi}{\bibitem}







\begin{center}

\begin{picture}(105,30)(0,38)
\Text(-50,40)[r]{(a):}
\Line(-40,55.75)(0,40.75)
\Line(-40,54.25)(0,39.25)
\Text(-20,57)[b]{$k-p$}
\ArrowLine(-21,48)(-19,47)
\ArrowLine(0,40)(-40,25)
\Text(-20,25)[t]{$-p$}
\Vertex(0,40){3}
\DashLine(0,40)(40,40){5}
\GOval(70,40)(15,30)(0){1}
\CArc(70,11)(43,47,133)
\DashLine(100,40)(140,40){5}
\Vertex(100,40){3}
\Vertex(40,40){3}
\Line(140,40.75)(180,55.75)
\Line(140,39.25)(180,54.25)
\ArrowLine(159,47)(161,48)
\Text(160,57)[b]{$k-p'$}
\ArrowLine(180,25)(140,40)
\Text(160,25)[t]{$-p'$}
\Vertex(140,40){3}
\end{picture}

\vspace{0.4in}

\begin{picture}(105,30)(0,38)
\Text(-50,40)[r]{(b):}
\Line(-40,55.75)(0,40.75)
\Line(-40,54.25)(0,39.25)
\ArrowLine(-21,48)(-19,47)
\ArrowLine(0,40)(-40,25)
\Vertex(0,40){3}
\GOval(37.5,40)(15,30)(0){1}
\CArc(37.5,11)(43,47,133)
\Vertex(7.5,40){3}
\Vertex(67.5,40){3}
\GOval(105,40)(15,30)(0){1}
\CArc(105,11)(43,47,133)
\Vertex(75,40){3}
\Vertex(135,40){3}
\Vertex(142.5,40){3}
\Line(142.5,40.75)(182.5,55.75)
\Line(142.5,39.25)(182.5,54.25)
\ArrowLine(161.5,47)(163.5,48)
\ArrowLine(182.5,25)(142.5,40)
\end{picture}

\vspace{0.4in}

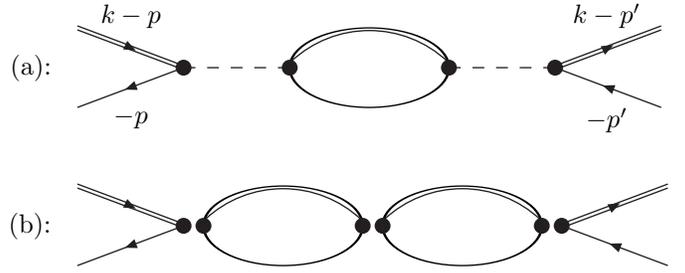
Fig.~1~~Heavy-light quark scattering in (a)Meson-quark\\
coupling picture, and (b)Quark-quark coupling picture.

\vspace{0.5in}

\begin{picture}(105,30)(0,38)
\Text(-20,40)[r]{(a):}
\DashLine(0,40)(40,40){5}
\ArrowLine(19,40)(21,40)
\Text(20,44)[b]{$\overline{\Lambda}v$}
\GOval(70,40)(15,30)(0){1}
\Photon(70,54.5)(105,70){2}{4.5}
\Vertex(70,54.5){2}
\Text(70,62)[b]{$\Gamma_\mu$}
\DashLine(100,40)(140,40){5}
\ArrowLine(119,40)(121,40)
\Text(120,44)[b]{$\overline{\Lambda}v'$}
\Vertex(100,40){3}
\Vertex(40,40){3}
\CArc(70,11)(43,47,133)
\ArrowLine(56,52)(54,53.5)
\ArrowLine(86,52)(84,50.5)
\ArrowLine(71,25)(69,25)
\Text(70,22)[t]{$-p$}
\end{picture}

\vspace{0.4in}

\begin{picture}(105,30)(0,38)
\Text(-20,40)[r]{(b):}
\DashLine(0,40)(40,40){5}
\ArrowLine(19,40)(21,40)
\Text(20,44)[b]{$\overline{\Lambda}v$}
\GOval(70,40)(15,30)(0){1}
\CArc(70,11)(43,47,133)
\Photon(100,40)(136,40){2}{4.5}
\Text(105,30)[t]{$\Gamma_\mu$}
\Vertex(40,40){3}
\Vertex(100,40){2}
\ArrowLine(69,54.5)(71,54.5)
\ArrowLine(71,25)(69,25)
\Text(70,22)[t]{$-p$}
\end{picture}

\vspace{.4in}

\begin{picture}(105,30)(0,38)
\Text(-20,40)[r]{(c):}
\DashLine(0,40)(40,40){5}
\ArrowLine(19,40)(21,40)
\Text(20,44)[b]{$\overline{\Lambda}v$}
\GOval(70,40)(15,30)(0){1}
\ArrowLine(69,54.5)(71,54.5)
\DashLine(100,40)(140,40){5}
\ArrowLine(119,40)(121,40)
\Text(120,44)[b]{$\overline{\Lambda}v$}
\Vertex(100,40){3}
\Vertex(40,40){3}
\CArc(70,11)(43,47,133)
\ArrowLine(54,27)(56,28.5)
\ArrowLine(84,27.3)(86,25.5)
\Text(70,25)[]{$\otimes$}
\Text(70,20)[t]{$A_\mu$}
\Text(88,25)[t]{$-p$}
\Text(52,25)[t]{$-p$}
\end{picture}

\vspace{0.4in}

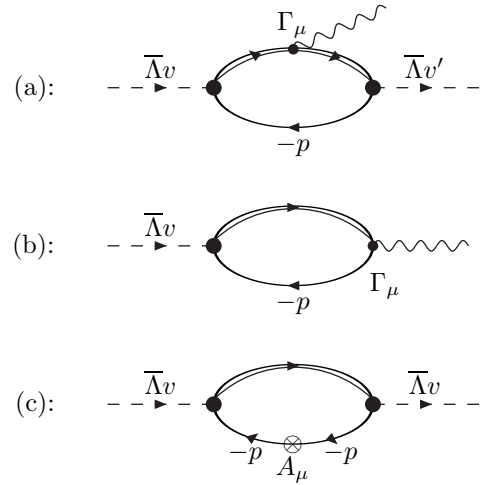
Fig.~2~~Feynman diagrams for (a)Isgur-Wise function,\\ 
(b)Decay constant, and (c)Axial vector coupling constant.

\vspace{0.5in}

\begin{picture}(145,30)(0,38)
\DashLine(0,40)(40,40){5}
\ArrowLine(19,40)(21,40)
\Text(20,44)[b]{$\overline{\Lambda}v$}
\GOval(70,40)(15,30)(0){1}
\Vertex(100,40){3}
\Vertex(40,40){3}
\CArc(70,11)(43,47,133)
\DashLine(100,40)(140,40){5}
\ArrowLine(119,40)(121,40)
\Text(120,44)[b]{$\overline{\Lambda}v'$}
\ArrowLine(69,54.5)(71,54.5)
\Gluon(70,25)(60,54){3}{4}
\ArrowLine(54,27)(56,28.5)
\ArrowLine(84,27.3)(86,25.5)
\Vertex(60,54){2} 
\Vertex(70,25){2} 
\Photon(80,53.5)(105,70){2}{4.5}
\Vertex(80,53.5){2}
\Text(60,64)[b]{${\cal O}({1\over m_Q})$}
\Text(85,50)[t]{$\Gamma_\mu$}
\end{picture}

\vspace{0.4in}

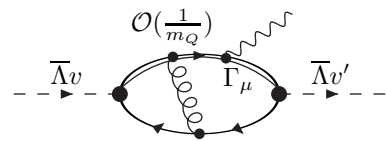
Fig.~3~~${\cal O}(1/m_Q)$ correction to Isgur-Wise function.
\end{center}



\end{document}